# Planned, delivered and variable RBE dose difference analysis for a patient cohort with base-of-tongue cancer treated with IMPT

Qianxia Wang[1, 2], Edgar Gelover Reyes[2], Alex Stanforth[2], William Andrew LePain[2], Haijian Chen[2], Mingyao Zhu[2], Katja M. Langen[2], William Stokes[2], Soumon Rudra[2], Mark McDonald[2], James Edward Bates[2], Stella Flampouri[2]

[1]Department of Radiation Oncology, University of Nebraska Omaha, 6001 Dodge St, Omaha, NE 68182, USA
[2] Department of Radiation Oncology and Winship Cancer Institute, Emory University, Atlanta, GA 30308, United States of America

**Background:**

To our knowledge, no tools have been installed in clinic for delivered and planned dose differences evaluation. This difference could be large for head and neck patients who suffer the most anatomy changes compared with other treatment sites due to long treatment courses and difficulty in eating.

At the same time, variable RBE dose is an increasing concern for proton therapy. The constant RBE 1.1 is widely applied in clinics, however, the real RBE is larger than 1.1 especially at the end of beam range. How they are different and what the influence on plan evaluation are an interesting topic to investigate.

**Purpose:** To develop a RayStation script-based tool for automatically calculating planned ($D_{planned}$), delivered ($D_{delivered}$) and variable RBE dose ($D_{RBE}$) and assessing dose difference using different quantities for patient cohorts, apply it to 47 patients with base-of-tongue cancer treated with IMPT and discuss clinical influence of the dose difference.

**Methods:** The $D_{planned}$ was calculated based on the TPCT and those QACTs with adaptive plans attached to them. The $D_{delivered}$ was accumulated doses calculated on the TPCT and all QACTs acquired during treatment with different weights. Both $D_{planned}$ and $D_{delivered}$ were with a constant RBE of 1.1. The calculation of $D_{RBE}$ was similar to $D_{delivered}$ but using McNamara model as the RBE. Quantities used to evaluate the dose difference include the gamma index passing rate ($P_{33}$), point-to-point dose difference passing rate ($P_3$), the difference in the mean dose ($\Delta\overline{D}$), three DVH indices ($\Delta D_{95}$, $\Delta D_{50}$, $\Delta D_{05}$) , and a selected NTCP model ($\Delta$NTCP).

**Results:** For $D_{planned}$ VS $D_{delivered}$, $\overline{P}_{33}$ was 58%, 53% and 56% for the oral cavity, esophagus and larynx respectively. The dose difference ($D_{planned}$–$D_{delivered}$) range are [-4.3, 3.9], [-3.1, 2.8], [-6.9, 2.7] and [-8.9, 5.9] in Gy for $\Delta\overline{D}$, $\Delta D_{95}$, $\Delta D_{50}$ and $\Delta D_{05}$ in the oral cavity; [-12.5,4.0] , [-3.9, 2.1], [-15.7, 2.6] and [-16.5, 8.5] in the esophagus; [-10.0, 5.7] , [-6.1, 2.9], [-9.9, 8.8] and [-14.2,8.1] in the larynx. The ΔNTCP is < 2.0%.

For $D_{RBE}$ VS $D_{delivered}$, $\overline{P}_{3}$ was 32%, 25% and 25% for the three OARs respectively. The dose difference ($D_{RBE}$–$D_{delivered}$) range are [1.1, 4.9], [0.0, 1.1], [0.1, 6.7] and [-0.4, 9.0] in Gy for $\Delta\overline{D}$, $\Delta D_{95}$, $\Delta D_{50}$ and $\Delta D_{05}$ in the oral cavity; [0.4, 6.7], [0.0, 3.0], [0.0, 7.4] and [2.1, 12.0] in the esophagus; [1.8, 6.5], [0.5, 3.1], [1.7, 8.1] and [-2.6, 7.9] in the larynx. The ΔNTCP is < 3.0%.

**Conclusion:** The tool efficiently calculated the three different doses and assessed dose discrepancies for the 47 base-of-tongue cancer patients. Low passing rates in $P_{33}$/$P_3$ and large deviations in DVH indices were observed in three OARs close to the primary or secondary targets. Large negative differences between $D_{planned}$ and $D_{delivered}$ and large positive differences between $D_{RBE}$ and $D_{delivered}$ were observed in $\overline{D}$, $\Delta D_{50}$ and $\Delta D_{05}$ for some patients. Special attention should be paid to these patients because their OARs may receive much more real doses beyond the constraints which cannot be detected by $D_{planned}$. used in clinic for plan evaluation. The selected NTCP model showed minimal variation across different doses.

## 1. INTRODUCTION

The most important concept in radiation therapy is dose which is used for plan optimization and evaluation, quality assurance (QA) verification and outcome analysis. The Monte Carlo algorithm[1] adapted to commercial treatment planning systems (TPS) improved the accuracy of dose calculation for proton therapy, especially for lung patients[2,3]. However, some elements such as stopping power uncertainty, patient setup, patient breath during treatment, and anatomic changes during treatment course can cause actual received dose to differ from the planned one. The influence of anatomic changes on doses can be monitored by routinely scanned quality assurance CT (QACT) during the treatment. The difference between doses based on the QACT and treatment planning CT (TPCT) determines if adaptive plan is required. This dose difference could be larger for head and neck patients who usually suffer the most anatomy changes compared with other treatment sites due to long treatment courses (6 - 7 weeks) and difficulty in eating.

Another challenge for proton therapy is the relative biological effectiveness (RBE). A constant RBE of 1.1 has been applied for proton therapy in clinics to conservatively estimate the biological dose in target area since the early days.[4] This approximation ignored that the RBE is a function of linear energy transfer (LET) and some other biological factors which results in underestimation of actual dose as large as 20%[5] in clinical practice. This simplification in RBE could cause irreversible injury to patients.[6] The RBE can be estimated with different models[7], including linear-quadratic cell survival model[8], empirical model[9–11] and Mechanism-inspired model[12–14]. McNamara model[11] is one of the empirical models developed based on all in vitro cell survival data published before. This model was used for variable RBE dose estimation in this study.

To our knowledge, no tools have been installed in the clinic for conveniently evaluating the delivered and planned dose differences due to anatomy changes during the treatment course. At the same time, very few clinical TPSs are enabled for variable RBE dose calculation. As far as we know, RayStation® is the only one that can estimate variable RBE dose distribution for patients. We developed a tool within RayStation scripting module to automatically calculate planned dose, delivered dose and variable RBE dose and assess dose differences with different quantities.

We applied this tool to 47 patients and quantitatively assessed dose difference with gamma index analysis[15] or point-to-point dose difference passing rate[16], the mean dose $\overline{D}$ and DVH[17] indices such as $D_{95}$, $D_{50}$, $D_{05}$ and a selected NTCP[18] model with the end point of moderate-to-severe xerostomia at 6 months after the radiation treatment. All the calculations and comparisons were performed within the tool developed with the RayStation script module that can be easily installed for regular clinics use.

## 2. MATERIALS AND METHODS

### 2.A. Patient cohort and treatment planning

We investigated 47 patients with base-of-tongue cancer treated with intensity modulated proton therapy (IMPT) from 2019 to 2023 at Emory Proton Therapy Center. Among these patients, 6 (13%) were female and 41 (87%) were male with diagnosis ages ranging 43 to 83 years old. These patients were treated bilaterally with a 5-field beam arrangement including two anterior oblique fields, two posterior oblique fields and one anterior-posterior field. The primary site prescribed dose and total fractions for these patients were 70 Gy/35 fx (85%), 69.96 Gy/33 fx (9%) or 66 Gy/30 fx (6%). All these patients were treated with Simultaneously Integrated Boost (SIB) plans generated on a TPCT scanned about 10 days before the first treatment fraction by a Monte Carlo-based commercial treatment planning system. Among these patients, 25 (53.2%) of them have 3 dose levels, 19 (40.4%) had 2 dose levels and 3 (6.4%) had 1 dose level. Adaptive plans were required for the remaining fractions if a patient's anatomical changes caused serious dosimetric consequence monitored by routinely scanned QACT during the treatment course. Three (6%) patients have 2 adaptive plans, 24 (51%) patients have 1 adaptive plan and the rest 20 patients (43%) had no adaptive plan.

| Condition | Description | Patient # | Percentage |
|---|---|---|---|
| Age (years) | Range | 43-83 | |

| | Median | | 70 | |
|---|---|---|---|---|
| Sex | Female | 6 | | 13% |
| | Male | 41 | | 87% |
| Dose to the primary site (Gy RBE) | 70/35 fx | 40 | | 85% |
| | 66.96/33 fx | 4 | | 9% |
| | 66/30 fx | 3 | | 6% |
| Dose levels (SIB) | 3 levels | 25 | | 53.2% |
| | 2 levels | 19 | | 40.4% |
| | 1 level | 3 | | 6.4% |
| Patients with adaptive plans | 0 | 20 | | 43% |
| | 1 | 24 | | 51% |
| | 2 | 3 | | 6% |

Table 1 Patient characteristics

**2.B. Dose calculation**

Three different doses for each patient were calculated in this study: the planned, delivered and variable RBE doses. The planned and delivered doses with 1.1 RBE were calculated by Monte Carlo in RayStation 10B and the variable RBE dose (McNamara model) were calculated in RayStation 11B-IonPG (research version) with radiobiological parameters $\alpha_x = 0.0499\ Gy^{-1}$, $\beta_x = 0.0238\ Gy^{-2}$ for normal tissues[19] and $\alpha_x = 0.22\ Gy^{-1}$, $\beta_x = 0.0272\ Gy^{-2}$ for targets[20].

For patients without any adaptive plans, the planned dose was calculated only on the TPCT. For patients with adaptive plan(s), the planned dose was the sum of the dose calculated on the TPCP using the original plan and the dose(s) calculated on the QACT(s) using the attached adaptive plan(s) with different weights. The weight for a specific plan was determined by the number of fractions delivered with this plan.

For the delivered and variable RBE doses, we assigned 1 for the dose based on TPCT because the TPCT was usually scanned 10 days before the treatment. The weight for a dose based on a QACT was determined by the number of fractions delivered close to the date when the QACT was scanned.

To enable the summation of doses from the TPCT and different QACTs, deformable registration was performed from each QACT to the TPCT and doses based on QACTs were mapped to the TPCT according to the registration.

**2.C. Gamma index analysis and point-to-point dose comparison**

Gamma index analysis was applied to compare the planned and delivered dose. For each patient, the delivered dose was set as the reference and 3%/3 mm was used as the criterion ($P_{33}$). When calculating the gamma index passing rate, we excluded voxels with doses below 5% of the maximum dose. In this study, we not only investigated the gamma index passing rate within the whole body but also within targets and 6 OARs.

The point-to-point dose comparison was used for the variable and constant RBE dose difference evaluation. Similar to the gamma index analysis, we also defined a passing rate criterion of 3% ($P_{3,}$ percentage of points with dose difference less than 3% of the maximum dose). Compared with the gamma index analysis used to evaluate the planned and delivered dose difference, the spatial uncertainty (3 mm) was removed because there was no spatial uncertainty between different RBE models.

**2.D. Average dose, DVH indices and an NTCP prediction model**

To evaluate the influence of dose difference on plan quality, we calculated four important plan evaluation parameters ($\overline{D}$, $D_{95}$, $D_{50}$ and $D_{05}$), and accessed their differences when comparing $D_{delivered}$ with $D_{planned}$ and $D_{delivered}$ with $D_{RBE}$ within 8 structures.

To evaluate the influence on toxicity probability prediction with planned/delivered/variable RBE dose, we selected an NTCP model of which the end point is moderate-to-severe xerostomia at 6 months after radiation treatment:

$NTCP = (1+e^{-LP})^{-1}$, where $LP = -1.507 + 0.052 \times D_{mean\ contralateral\ parotid\ gland}$

$+ 0.525 \times$ minor $XER_{baseline}$ (1 if $XER_{baseline}$ is minor, 0 if not)

$+ 1.482 \times$ moderate-to-severe $XER_{baseline}$ (1 if $XER_{baseline}$ is moderate or severe, 0 if not).

All the patients selected in this study was treated bilaterally. So instead of using mean dose within the contralateral parotid, we used mean dose with left and right parotids.

## 3. RESULTS

### 3.A. The planned dose VS the delivered dose

The planned and delivered dose difference was analyzed by gamma index analysis with a criterion of 3%/3 mm (Table 2 and Figure 1 (a)). Among all structures investigated in this study, the target had the best passing rate ($\overline{P}_{33} = 92\%$). The mean passing rate over the 47 patients for the whole body were 74% with an uncertainty of 7%.

The rest of the OARs (the left and right parotid, oral cavity, mandible bone, esophagus, and larynx) had low mean passing rates and large standard deviations. Compared with the other four OARs, the two parotids have relatively higher passing rates with an average value (uncertainty) of 83% (18%) for the left parotid and 84% (18%) for the right parotid respectively. The oral cavity, esophagus and larynx have the lowest passing rates with an average value (uncertainty) of 58% (14%), 53% (22%) and 56% (15%), respectively.

The planned and delivered dose difference (planned dose – delivered dose) was also assessed by mean dose and three different DVH indices (Table 2 and Figure 2(a)) which were used for plan evaluation or outcome analysis in clinics. Among all structures investigated in this study, the differences are the largest in $D_{05}$ and the smallest in $D_{95}$ compared to other DVH indices or mean dose. Similar to the $P_{33}$, the mean dose and DVH indices agreed well in structures of body and target. Larger deviations were observed for the rest of the 6 structures. The oral cavity, esophagus and larynx have considerable deviation in $\overline{D}$ and DVH indices.

Among the 47 patients, ranges were [-3.7, 2.7], [-1.3, 0.7], [-5.1, 2.9] and [-5.4, 6.6] in Gy for $\Delta\overline{D}$, $\Delta D_{95}$, $\Delta D_{50}$, and $\Delta D_{05}$ for parotids. For the mandible bone, ranges were [-3.5, 1.2], [-1.7, 1.1], [-5.7, 1.8] and [-5.4, 3.0] in Gy for $\Delta\overline{D}$, $\Delta D_{05}$, $\Delta D_{50}$, and $\Delta D_{95}$ respectively. For the three OARs (the oral cavity, esophagus and larynx), ranges for $\Delta\overline{D}$, $\Delta D_{05}$, $\Delta D_{50}$ and $\Delta D_{95}$ were [-4.3, 3.9], [-3.1, 2.8], [-6.9, 2.7] and [-8.9, 5.9] in Gy for the oral cavity; [-12.5, 4.0], [-3.9, 2.1], [-15.7, 2.6] and [-16.5, 8.5] for the esophagus; [-10.0, 5.7], [-6.1, 2.9], [-9.9, 8.8] and [-14.2, 8.1] for the larynx. In all structures, the median values of $\Delta\overline{D}$, $\Delta D_{05}$, $\Delta D_{50}$ and $\Delta D_{95}$ were small (within ± 1.8 Gy).

We also tested the influence of dose difference between the planned and delivered doses on one NTCP model. The ΔNTCP (%) for the 47 patients is between -1.9% and 1.1% with a median value of -0.3%.

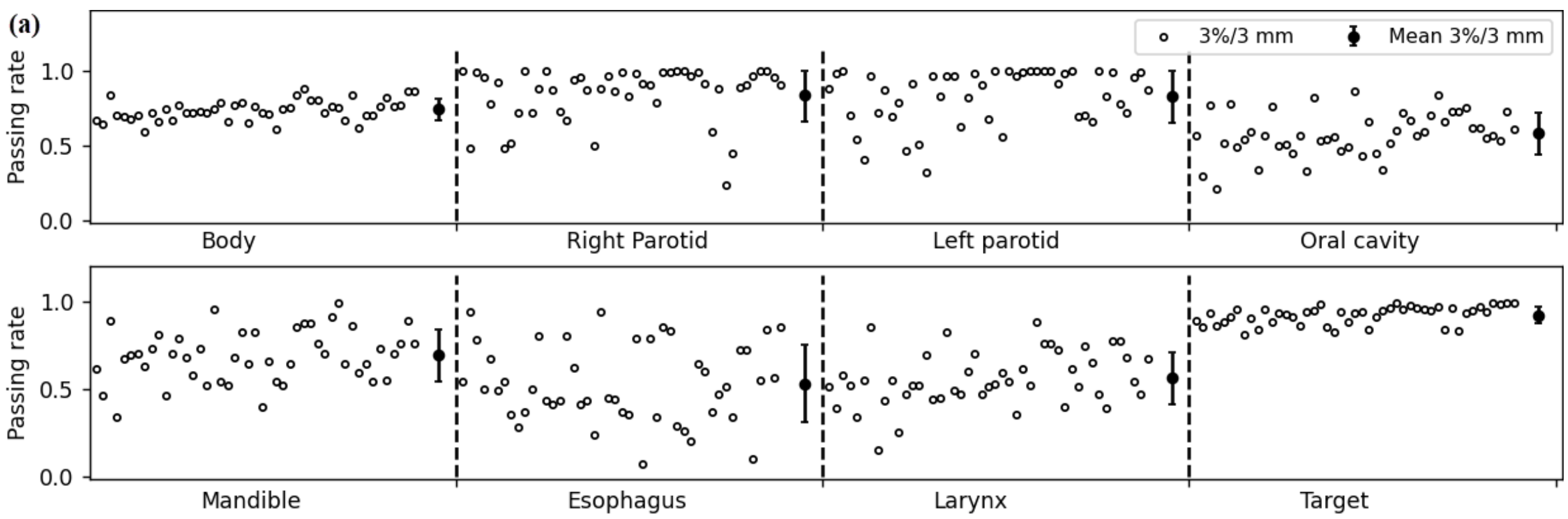

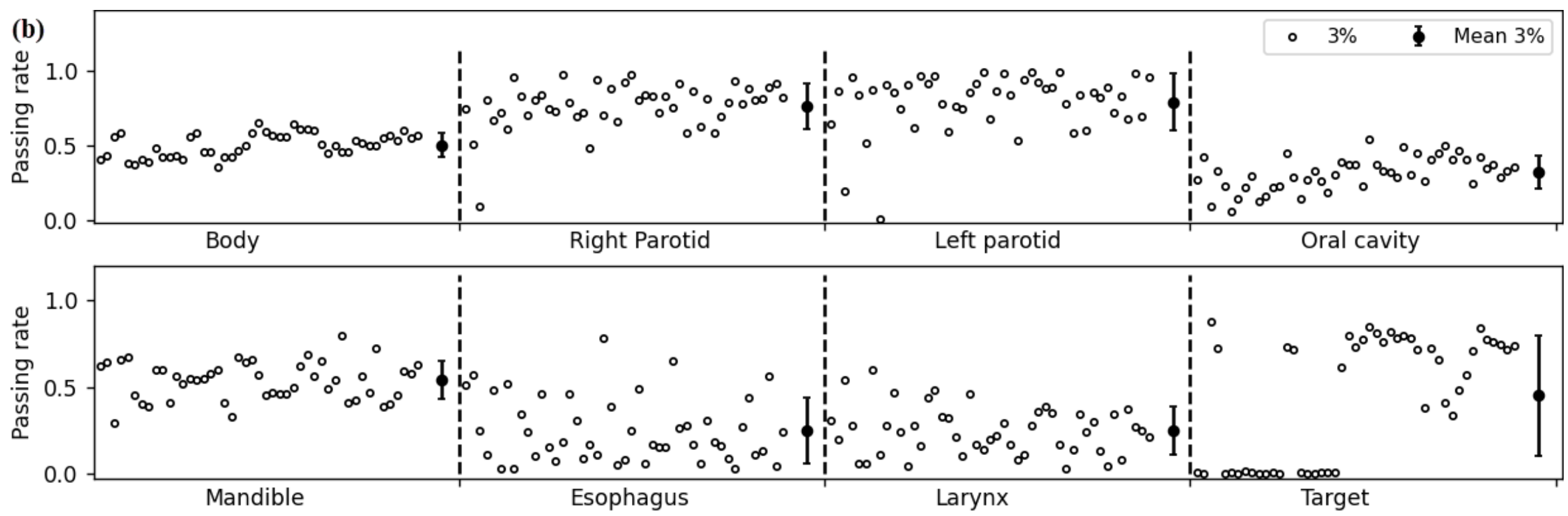

Figure 1. (a) The planned dose ($D_{planned}$) VS the delivered dose ($D_{delievered}$): Gamma index passing rate for different structures with the criterion of 3%/3 mm; (b) The constant RBE dose ($D_{delievered}$) VS the variable RBE dose ($D_{RBE}$): point dose difference passing rate for different structures with the criterion of 3%. Markers without a bar were passing rates for individual patients. Markers with a bar were the mean passing rates of all patients and standard deviations.

| Doses compared | OAR | $\overline{P}_{33}/\overline{P}_3 \pm$ SD | $\Delta\overline{D}$ (Gy) median [range] | $\Delta D_{95}$(Gy) median [range] | $\Delta D_{50}$ (Gy) median [range] | $\Delta D_{05}$ (Gy) median [range] | ΔNTCP (%) median [range] |
|---|---|---|---|---|---|---|---|
| (a) Planned VS Delivered dose | Body | 0.74 ± 0.07 | -0.1 [-0.4, 0.2] | 0.0 [0.0, 0.0] | 0.0 [0.0, 0.0] | -0.1 [-1.2, 2.2] | -0.3 [-1.9, 1.1] |
| | Parotid (R) | 0.84 ± 0.18 | -0.5 [-3.7, 2.7] | -0.1 [-1.1, 0.7] | -0.5 [-5.1, 2.9] | 0.0 [-3.7, 6.6] | |
| | Parotid (L) | 0.83 ± 0.18 | 0.0 [-3.2, 1.8] | 0.0 [-1.3, 0.3] | -0.1 [-4.8, 2.2] | 0.3 [-5.4, 3.5] | |
| | Oral Cavity | 0.58 ± 0.14 | -1.4 [-4.3, 3.9] | 0.0 [-3.1, 2.8] | -1.8 [-6.9, 2.7] | 0.1 [-8.9, 5.9] | |
| | Mandible | 0.69 ± 0.15 | -1.0 [-3.5, 1.2] | 0.0 [-1.7, 1.1] | -1.5 [-5.7, 1.8] | -0.7 [-5.4, 3.0] | |
| | Esophagus | 0.53 ± 0.22 | -0.8 [-12.5, 4.0] | 0.0 [-3.9, 2.1] | -1.0 [-15.7, 2.6] | -0.7 [-16.5, 8.5] | |
| | Larynx | 0.56 ± 0.15 | -0.8 [-10.0, 5.7] | -0.8 [-6.1, 2.9] | -1.5 [-9.9, 8.8] | 0.2 [-14.2, 8.1] | |
| | Target | 0.92 ± 0.05 | 0.1 [-0.6, 0.7] | 0.6 [0.0, 1.8] | 0.1 [-0.5,0.6] | 0.0 [-2.4, 0.7] | |
| (b) Variable VS | Body | 0.50 ± 0.08 | 0.5 [0.4, 0.9] | 0.0 [0.0, 0.0] | 0.0 [0.0, 0.1] | 4.5 [1.8, 5.8] | 1.2 [0.5, 2.7] |
| | Parotid (R) | 0.76 ± 0.15 | 1.1 [-0.4, 6.5] | 0.1 [-0.6, 1.6] | 0.9 [-1.2, 7.9] | 2.9 [-1.5, 8.4] | |
| | Parotid (L) | 0.79 ± 0.19 | 0.9 [0.3, 4.9] | 0.1 [-0.1, 3.9] | 0.8 [-0.1, 5.1] | 2.4 [-1.8, 5.6] | |
| | Oral Cavity | 0.32 ± 0.11 | 2.6 [1.1, 4.9] | 0.1 [0.0, 1.1] | 2.6 [0.1, 6.7] | 4.9 [-0.4, 9.0] | |

| constant RBE dose | Mandible | 0.54 ± 0.11 | 1.7 [0.9, 3.7] | 0.1 [0.0, 1.0] | 1.4 [0.3, 3.8] | 4.5 [2.2, 7.1] | |
|---|---|---|---|---|---|---|---|
| | Esophagus | 0.25 ± 0.19 | 2.7 [0.4, 6.7] | 0.0 [0.0, 3.0] | 2.3 [0.0, 7.4] | 6.7 [2.1, 12.] | |
| | Larynx | 0.25 ± 0.14 | 3.8 [1.8, 6.5] | 1.3 [0.5, 3.1] | 4.0 [1.7, 8.1] | 5.9 [-2.6, 7.9] | |
| | Target | 0.45 ± 0.35 | 1.0 [-1.0, 6.7] | 0.8 [-2.0, 5.8] | 1.0 [-1.0, 6.6] | 1.6 [-0.8, 7.9] | |

Table 2. (a) The planned dose ($D_{planned}$) VS the delivered dose ($D_{delivered}$): mean Gamma index passing rate and their standard deviation with the criterion of 3%/3 mm ($P_{33}$); median and range of difference in the mean dose ($\overline{D}$), three DVH indices ($D_{95,}$ $D_{50}$ and $D_{05}$) for different structures and NTCP between the planned and delivered dose (the planned dose – the delivered dose); (b) The constant RBE dose ($D_{delivered}$) VS the variable RBE dose ($D_{RBE}$): mean point-to-point dose difference passing rate and their standard deviation with the criterion of 3% ($P_3$), median and range of difference in the mean dose ($\overline{D}$), three DVH indices ($D_{95,}$ $D_{50}$ and $D_{05}$) for different structures and NTCP between the planned and delivered dose (the variable RBE dose – the constant RBE dose).

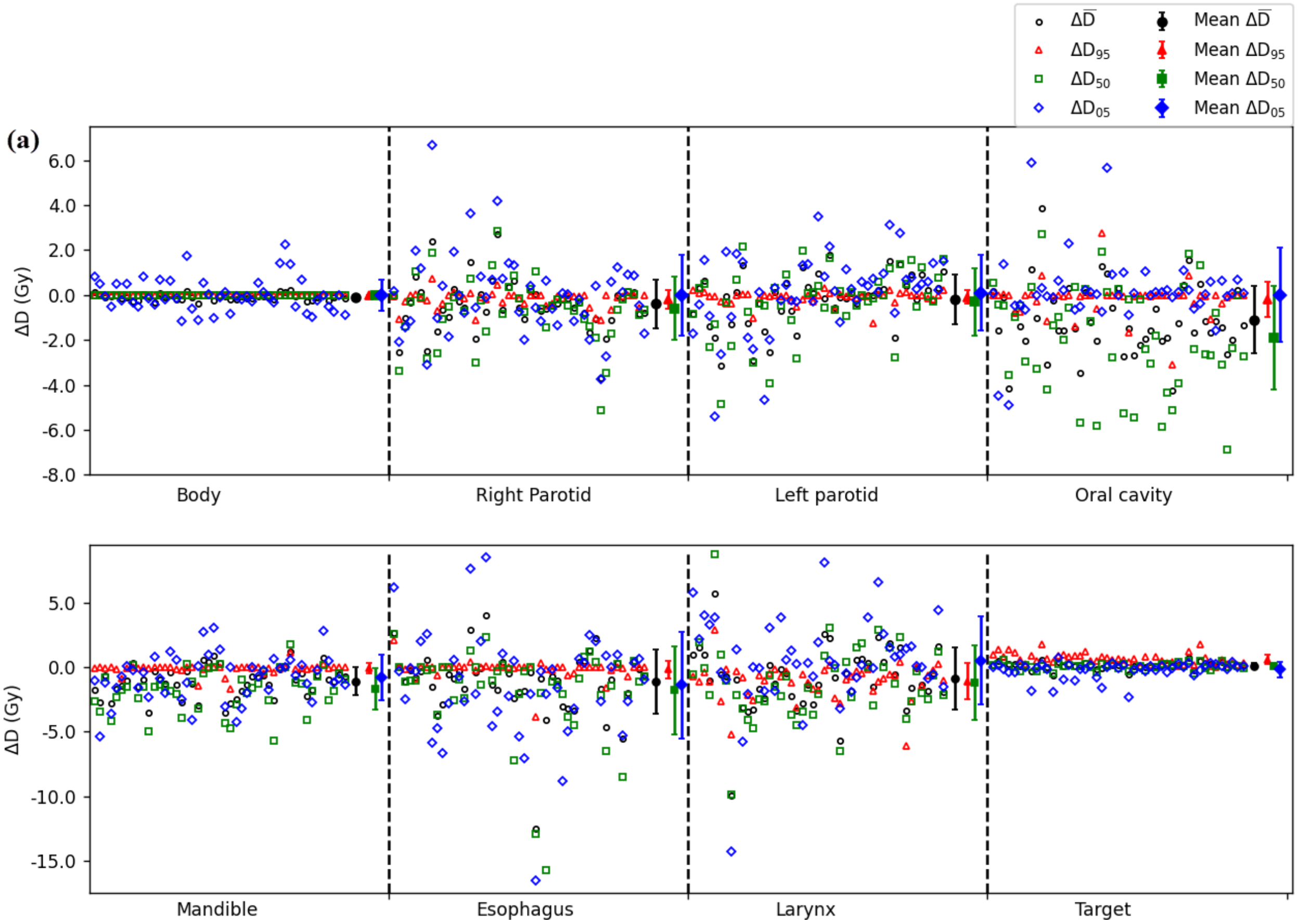

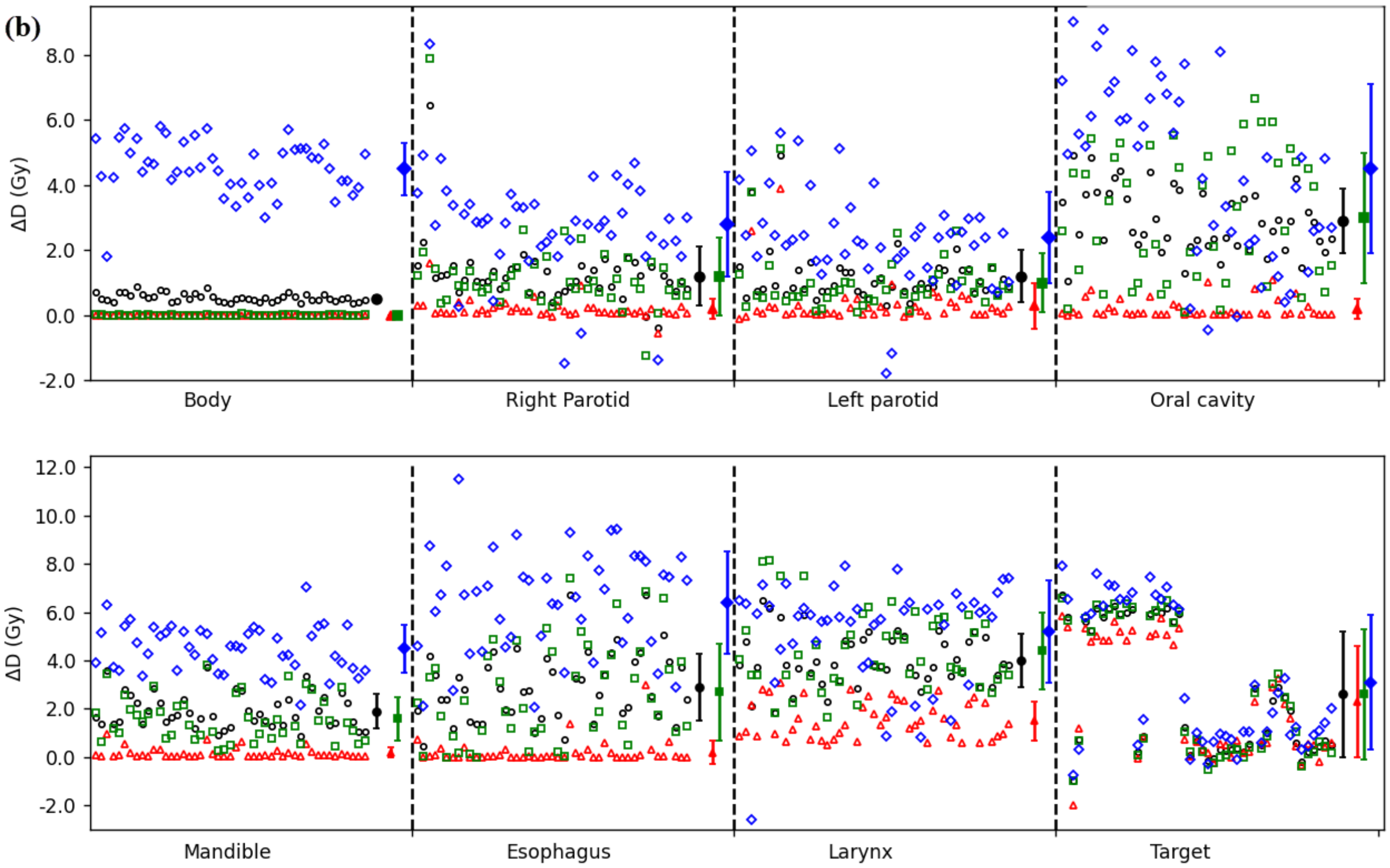

Figure 2. (a) The planned dose ($D_{planned}$) VS the delivered dose ($D_{delivered}$): difference in the mean dose ($\overline{D}$), three DVH indices ($D_{95}$, $D_{50}$ and $D_{05}$) (the planned dose – the delivered dose); (b) The constant RBE dose ($D_{delivered}$) VS the variable RBE dose ($D_{RBE}$): difference in the mean dose ($\overline{D}$), three DVH indices ($D_{95}$, $D_{50}$ and $D_{05}$) (the variable RBE dose – the constant RBE dose). Markers without a bar were the values of individual patients. Markers with a bar were the mean values of all patients and standard deviations.

### 3.B. The variable RBE dose VS the constant RBE dose

The variable and constant RBE dose difference was analyzed by point-to-point dose difference passing rate (Table 2 and Figure 1(b)) instead of gamma index analysis. Within the whole body, 50% of voxels on average for the 47 patients are with dose difference less than 3% of maximum dose. The percentage is 45% on average for the target and 54% for mandible bone. For parotid, there are more voxels with dose difference less than 3% (76% for the right parotid and 79% for the left parotid). The oral cavity, esophagus and larynx have much fewer voxels with dose difference less than 3%. The passing rate is 32% for the oral cavity, 25% for the esophagus and 25% for the larynx.

The difference between the variable and constant RBE doses (variable RBE dose - constant RBE dose) were investigated by mean dose and three DVH indices (Table 2 and Figure 2(b)). Generally speaking, variable RBE doses were higher than the constant RBE doses. $D_{95}$ has the largest and $D_{05}$ has the smallest deviation between the two doses compared to the mean dose or other DVH indices. Among different structures, the oral cavity, esophagus and larynx have the largest deviation in the mean dose and DVH indices.

Ranges of $\Delta\overline{D}$, $\Delta D_{95}$, $\Delta D_{50}$ and $\Delta D_{05}$ were [1.1, 4.9], [0.0, 1.1], [0.1, 6.7] and [-0.4, 9.0] in Gy for the oral cavity, [0.4, 6.7], [0.0, 3.0], [0.0, 7.4] and [2.1, 12.0] for esophagus; [1.8, 6.5], [0.5, 3.1], [1.7, 8.1] and [-2.6, 7.9] for larynx. Median values for $\Delta\overline{D}$ and $\Delta D_{50}$ were 2.3 Gy – 4.0 Gy for the three OARs. Median values for $\Delta D_{95}$ were ≤ 1.3 Gy for the three OARs. And median values for $\Delta D_{05}$ were 4.9 Gy, 6.7 Gy and 5.9 Gy respectively for the three OARs. Differences in $\overline{D}$, $D_{95}$, $D_{50}$ and $D_{05}$ had a range of [-0.4, 6.5], [-0.6, 3.9], [-1.2, 7.9] and [-1.8, 8.4] in Gy respectively within parotids. Ranges of the four quantities were [0.9, 3.7], [0.0, 1.0], [0.3, 3.8] and [2.2, 7.1] in Gy within mandible bone. Median values were ≤ 1.7 Gy for $\Delta\overline{D}$, $\Delta D_{95}$ and $\Delta D_{50}$, and 2.4 Gy – 4.5 Gy for $\Delta D_{05}$ for parotids and mandible bone. For the target, differences in $\overline{D}$, $D_{95}$, $D_{50}$ and $D_{05}$ were with small median values ( ≤1.6 Gy) and with a wide range of [-1.0, 7.9] in Gy.

No significant difference was observed in the selected NTCP model by using the variable and constant RBE doses. The median value was 1.2% and the range was [0.5%, 2.7%].

## 4. DISCUSSION

To our knowledge, there is no clinical tool installed for delivered dose calculation and quantitative dose difference evaluation between planned and delivered doses, constant RBE and variable RBE doses. We developed a tool within RayStation script module that allows us to conveniently calculate delivered/planned/variable RBE dose for a cohort of patients and evaluate dose difference between each pair of them. In this study, we applied this tool to 47 patient cohort with base-of-tongue cancer treat with IMPT.

Gamma index analysis was performed with a criterion of 3%/3 mm to analyze the difference between the planned and delivered dose at voxel level. Passing rates were calculated for 8 different structures which were body, left and right parotids, oral cavity, mandible bone, esophagus, larynx and targets. The spinal cord and brainstem were far away from the target, so they received low doses (some patients received dose less than 5% of maximum global dose). The two OARs were not included in this study. Passing rates of the body reflected the general dose deviation between the planned and delivered doses. Passing rates in targets were high for all patients. Because an adaptive plan was generated based on the newly acquired QACT if target dose coverage dropped, and the adaptive plan was included in the planned dose. The oral cavity, esophagus and larynx had lowest passing rates compared with other OARs because they were close to the primary or secondary targets and suffered more anatomic changes.

The difference in mean dose ($\overline{D}$) and 3 DVH indices ($D_{95}$, $D_{50}$ and $D_{05}$) were calculated to evaluate the influence on plan caused by dose differences. The deviation was small in target and large in three OARs close to targets, which was consistent with the trends of gamma index analysis.

The $\overline{D}$ and $D_{50}$ of the parotid, $\overline{D}$ of the oral cavity, esophagus and larynx, $D_{0.03cc}$ and $D_{42}$ of the mandible bone are quantities used for plan evaluation. Negative values for the difference in these quantities should be paid more attention because the planned dose is used for plan evaluation in clinic and negative difference means OARs' doses were underestimated. Among the 47 patients, $\overline{D}$ and $D_{50}$ calculated from the planned dose for parotids can be 3.7 Gy and 5.1 Gy lower than the delivered dose respectively. The $\overline{D}$ of the oral cavity, esophagus and larynx calculated from the planned dose could be as low as 4.3 Gy, 12.5 Gy and 10 Gy than that from the delivered dose. $D_{0.03cc}$ and $D_{42}$ for mandible bone were not calculated in this study, we used $D_{05}$ and $D_{50}$ to approximately represent the two values. The $\Delta D_{05}$ and $\Delta D_{50}$ of the mandible bone for the 47 patients can be as low as about -5.7 Gy and -5.4 Gy, respectively.

According to the analysis, considerable patients have lower DVH indices and mean dose calculated by planned dose than delivered dose. It means dose constraints for some OARs were met according to the planned dose but the real received dose may be beyond their dose limit. For those patients, it is dangerous to use planned dose for plan evaluation. A clinical tool that enables delivered dose calculation is pressing need.

Instead of using gamma index analysis to evaluate the difference between the constant and variable RBE doses, difference of each point (point-to-point dose difference) were calculated because there was no spatial uncertainty between the two sets of doses. Similarly, we observed that the oral cavity, esophagus and larynx have the lowest passing rates. The variable RBE doses were higher than the constant RBE doses especially at the end of the beam range. The oral cavity, esophagus and larynx were near the primary or secondary targets which were close to the end of beam field range and had largest dose deviations.

The mean dose and three DVH indices calculated from variable RBE doses were usually higher than those calculated from the constant RBE doses, which means the values used in clinic were underestimated. However, it is still immature to implement variable RBE in clinic for plan optimization due to the uncertainty in biological coefficients and how RBE varies according to dose, LET and biological coefficients. Besides, dose constraints for OARs used in clinic were from historical experience. For patient outcome evaluation, it is helpful to include the variable RBE doses to verify if the current OAR constraints work or need to be modified.

The NTCP model selected in this study does not show much difference (< 3%) when using planned, delivered or variable RBE doses for toxicity probability calculation. This model is a function of mean contralateral parotid dose which does not change much whether using planned, delivered or variable RBE methods to evaluate the dose distribution.

## CONFLICT OF INTEREST

We have no conflicts of interest to disclose.